# HYDRODYNAMIC EVOLUTION OF CHIRALLY SYMMETRIC NUCLEAR MATTER

F. S. NAVARRA,

Instituto de Física da Universidade de São Paulo,
Caixa Postal 20516, 01498-970 São Paulo, Brazil

U. ORNIK

Gesellschaft für Schwerionenforschung, Planckstr.1, 6100 Darmstadt, Germany

**Abstract:** The equation of state of the linear sigma model in the mean field approximation is used as input in a relativistic hydrodynamical numerical routine. Longitudinal and transverse energy distributions are calculated and compared with those obtained from the QHD-I equation of state.

## INTRODUCTION

The description of the hadronic matter equation of state based on hadronic degrees of freedom is very attractive. First of all because hadrons are the particles observed experimentally and are also the most efficient variables at low densities and temperatures. Second because extreme conditions can be extrapolated from calibrations made between hadronic calculations and observed hadron-hadron scattering and empirical nuclear properties. Finally because an accurate hadronic description is required to isolate and identify true signatures of the QCD behavior in nuclear matter.

The so far most successful field theoretical model for nuclear matter is QHD-I or Walecka's model. In an early work [1] we have applied it to the description of relativistic heavy ion collisions with the conclusion that the equation of state of QHD-I is too hard, producing too much pressure and too less entropy. One could in principle make changes in the lagrangian to correct these problems but here we take a different approach. Since chiral symmetry is a symmetry of QCD and it is also very important in ordinary hadronic physics (e.g. pion-nucleon interactions) it becomes very interesting to investigate how it can affect the thermodynamical properties of our hot and dense nuclear matter.

In this work we use the equation of state of the linear sigma model derived previously [2] within the mean field approximation. The version of the sigma model considered here does not give a good description of the nuclear matter ground state properties. As it is well known [3] with this model is difficult to saturate the nuclear matter at the correct density with the correct binding energy. This can be achieved at the mean field level by the introduction of a term with couples the scalar and the vector fields. This new term will however create new problems for the correct description of finite nuclei [3].

Chiral symmetry has two thermodynamical consequences. The first one comes from the fact that it fixes the nucleon-scalar meson coupling to $g = g_\pi$ . As it was already seen this model contains a phase transition to a phase of massless nucleons. One of the order parameters of this transition is the nucleon effective mass . It is determined for a given value of the temperature and chemical potential by solving the so called gap equation. We can then define a line in the $T - \mu$ plane along which a rapid change in the effective mass takes place. This line is called here the phase-boundary. It is strongly determined by the value of the scalar coupling. It is now fixed by chiral symmetry in opposition to other effective models where it is just a free parameter. The other important effect induced by chiral symmetry is that the chiral gap equation admits always $M^* = 0$. This will make the phase transition to happen faster and immediately after it nucleons will become massless and will constitute a gas of massless particles having therefore a larger pressure. How these features (which are very insensitive to details of the specific version of the chiral lagrangian) manifest themselves in the hydrodynamical evolution of hot and dense chiral matter will be discussed below.

Our starting point is the standard linear sigma model lagrangian density

$$\mathcal{L} = \overline{\Psi}[\gamma_\mu(i\partial^\mu - g_v V^\mu) - g(\sigma + i\gamma_5 \tau \cdot \pi)]\Psi + \frac{1}{2}(\partial_\mu \sigma \partial^\mu \sigma - \partial_\mu \pi \cdot \partial^\mu \pi)$$
$$-\frac{1}{4}F_{\mu\nu}F^{\mu\nu} + \frac{1}{2}m_v^2 V_\mu V^\mu - \frac{1}{4}\lambda(\sigma^2 + \pi^2 - u^2)^2 , \qquad (1)$$

where $F_{\mu\nu} = \partial_\mu V_\nu - \partial_\nu V_\mu$. Here $\Psi$, V, $\sigma$ and $\pi$ are respectively the nucleon, vector meson, scalar meson and pion field. $u$ is the vacuum expectation value of the sigma field which gives mass to the nucleon. In this version of the sigma model there is no symmetry breaking term and the symmetry is realized in the Nambu-Goldstone mode.

Repeating the steps presented in ref.[1] and [2] one can derive the equation of state and compare it to the QHD-I one. Both of them present the "massless nucleon" phase transition at high temperatures. For the standard values of the parameters $C_s^2 = g_s^2(M^2/m_s^2) = 357.4$, and $C_v^2 = g_v^2(M^2/m_v^2) = 273.8$ the critical temperature (at zero density) for QHD-I is 186 MeV whereas it is 145 MeV for the sigma model. The same happens at zero temperature: the transition occurs at $\mu = 0.4 GeV$ for QHD-I and at $\mu = 0.24 GeV$ for the sigma model.

## THE HYDRODYNAMIC EVOLUTION

In the hydrodynamical picture of a heavy-ion reaction one can divide the evolution of the system into three stages: first there is a compression and thermalization of nuclear matter, then this highly excited fireball begins to expand according to the laws of relativistic hydrodynamics and finally the system decouples and particles are emitted. Here we use Landau-type initial conditions. The system is assumed to be partially stopped and its energy is deposited homogeneously in a cylindrical initial volume, the longitudinal size of which is Lorentz-contracted. The initial size, energy and baryon number of the fireball are all the initial conditions which are required for the hydrodynamical expansion. In order to know any other thermodynamical quantity the EOS is needed.

After specifying the initial conditions the nuclear matter is treated as an ideal relativistic fluid obeying the laws of hydrodynamics which are written in a brief form as follows

$$\partial_\mu T^{\mu\nu} = 0 , \qquad (2)$$

$$\partial_\mu B^\mu = 0 , \qquad (3)$$

$$T^{\mu\nu} = (\mathcal{E} + P)u^\mu u^\nu - P g^{\mu\nu} , \qquad (4)$$

$$B^\mu = \rho_B u^\mu , \qquad (5)$$

where $P$, $\mathcal{E}$ and $\rho_B$ are the pressure, energy density and baryon density defined in the last section. $u^\mu$ is the four-velocity and $g^{\mu\nu}$ is the metric tensor. This coupled system of partial differential equations, together with the relation between $\mathcal{E}$ and $P$ given by the EOS, is solved numerically in 3+1 dimensions by the program HYLANDER. The hydrodynamical description ends when a critical "freeze-out" temperature $T_f$ or density $\rho_f$ is reached.

## NUMERICAL RESULTS

Integrating eq. (2) and eq. (3) we obtain energy density, temperature, baryon density, sound velocity and other termodynamic quantities as a function of the longitudinal coordinate x and transverse coordinate r for successive time steps. In a realistic detailed calculation great care should be taken of the "freeze-out" and comparison with experimental data would be possible. This will be done elsewhere. Now we concentrate on the qualitative aspects of the hydrodynamical evolution of a system which is governed by chiral symmetry and compare it to a typical "non-chiral one".

We will consider a collision between sulphur-sulphur $^{32}S + ^{32}S$ at the center of mass energy $\sqrt{s} = 19 AGeV$. We assume that all nucleons are captured in the collision volume

("full stopping") but have some initial velocity distribution at the beginning of the expansion. The whole energy contained in the fluid is approximately 600 GeV. Half of it is in the form of baryon kinetic energy and the other half is "thermal energy" (carried by the mesonic fields). This energy is uniformly distributed along a cylindrical volume of radius 3 fm and length 1.5 fm. At t=0 fm the energy density is then $7 GeV/fm^3$, the baryon number density is $0.6 fm^-3$ and the initial baryon velocity distribution is $v = const \cdot x$. Since the main goal of this work is not to obtain an accurate description of experimental data we do not further elaborate these initial conditions. We just would like to remark that they are not very different from others which result from very detailed calculations concerning the initial stage of these collisions. We expect therefore that the numbers coming from our simulations are consistent with CERN-SPS experiments.

In Fig. 1a we present the longitudinal energy distributions of the fluid at times t=0,1,2,3 and 4 fm .

The solid lines are obtained with QHD-I equation of state (EOS) and the dashed lines represent the evolution of a fluid obbeying the sigma model EOS. Comparing the evolution of solid and dashed lines at equal times we can see that they deviate from each other, dashed lines moving faster downwards and occupying faster the large x region. This means that "the chiral fluid" moves and cools faster.

In Fig. 1b we show the evolution of the transverse energy distribution. We observe the same as in fig.1 but the effect is stronger. At t=5 fm there is a difference of one order of magnitude in the energy density for small values of r.

**CONCLUSIONS**

This work follows the same line as ref.[3]. There, the authors, looking for a better field theoretical description of the ground state of nuclear matter try to incorporate chiral symmetry in it. For that purpose they examine the standard linear sigma model and also some attempts to improve it. They try to describe both nuclear matter and finite nuclei. They come to the conclusion that the sigma model and its variants do not seem to give a reasonable description of data. Here we play the same game trying now to incorporate chiral symmetry in nucleus-nucleus reactions. Our results indicate that the sigma model lives longer than QHD-I in the high pressure phase and expands faster. It does not look promising in what concerns an accurate description of data.

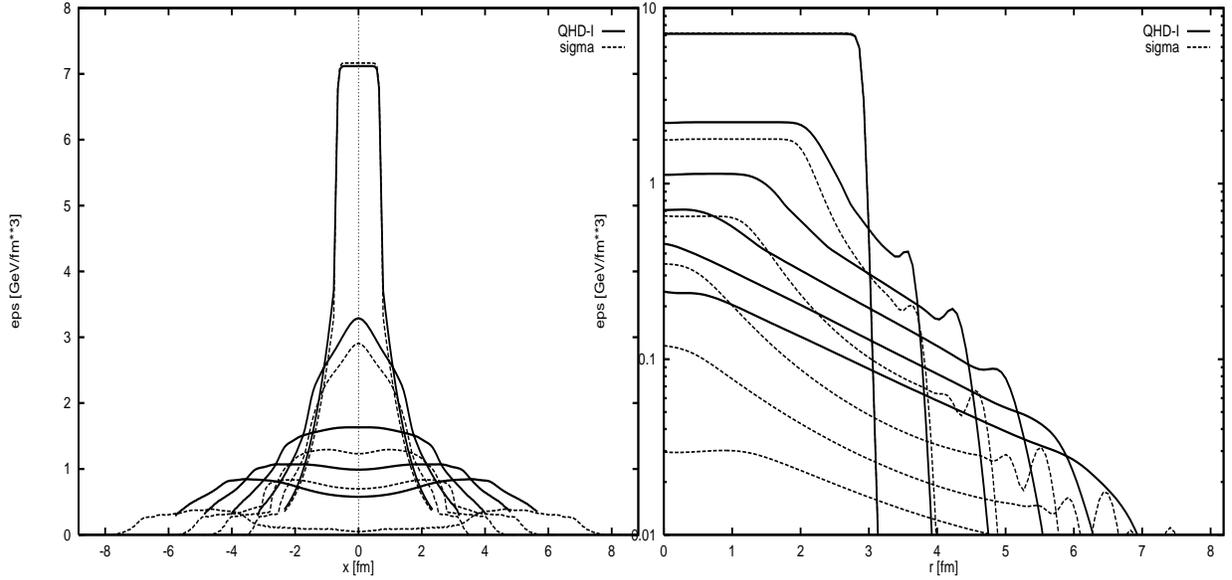

Figure 1: a)Longitudinal energy distributions at different time steps. b)Transverse energy distributions at different time steps.

## ACKNOWLEDGEMENTS

This work was partially supported by FAPESP – Brazil, CNPq – Brazil , DAAD (Germany) and GSI-Darmstadt.